\documentclass[prl,aps,reprint,preprintnumbers,amsmath,amssymb,superscriptaddress]{revtex4-1}

\usepackage{dsfont}
\usepackage{graphicx}
\usepackage{subfig}

\usepackage{xcolor}

\begin{document}

\title{The Entropy Shift and Extremality Shift at Zero Temperature}
\author{Brian McPeak}
\email{brian.mcpeak@df.unipi.it}
\affiliation{Department of Physics, University of Pisa and INFN, \\Largo Pontecorvo 3, I-56127 Pisa, Italy}

\begin{abstract}
    There is a well-established connection between the higher-derivative corrections to the black hole entropy and the black hole extremality bound. The particular combination of EFT coefficients $c_i$ that controls the mass shift at fixed charge and temperature also controls the entropy shift at fixed charge and mass in the limit where the mass approaches the \textit{uncorrected} value for the extremal mass. In this note, we use the classical entropy function formalism to examine the entropy corrections at exactly zero temperature, or at the \textit{corrected} value for the extremal mass. We find that the zero-temperature entropy shift (a) is unrelated to the mass shift, (b) is $\mathcal{O}(c_i)$ in the EFT coefficients, rather than $\mathcal{O}(\sqrt{c_i})$, as is the constant-mass entropy shift, and (c) is negative in the example of the EFT arising at low energies from QED plus gravity.
\end{abstract}

\maketitle
\flushbottom

\section{Introduction}
\label{sec:intro}

Gravity is well-described by General Relativity at low energies, but it is expected that the full quantum theory will introduce higher-derivative corrections to the Einstein-Hilbert action. There are certain situations where the absolute sign of these corrections can have a large effect. In field theory, for instance, corrections with a certain sign can lead to superluminal propagation on some backgrounds, thereby disallowing such corrections on causality grounds \cite{Adams:2006sv}. In gravity, the corrections can affect the extremality bound of black holes. As first pointed out in \cite{Kats:2006xp}, this is relevant for the Weak Gravity Conjecture (WGC) \cite{Arkani-Hamed:2006emk}, which roughly requires the existence of a superextremal state. The WGC can be satisfied by the black hole spectrum alone provided a certain combination of higher-derivative coefficients is positive. This means that the extremal mass decreases for black holes with a fixed charge.

Recently there has been a lot of progress on understanding if the required inequality does indeed hold using diverse arguments including consistency of the $S$-matrix \cite{Cheung:2014ega, Hamada:2018dde, Bellazzini:2019xts, Alberte:2020bdz}, RG running \cite{Charles:2019qqt, Jones:2019nev, Arkani-Hamed:2021ajd}, modular invariance \cite{Aalsma:2019ryi}, the $c$-theorem \cite{Aalsma:2020duv}, and investigating various specific examples \cite{Cano:2019oma, Cano:2019ycn, Cano:2021nzo} and consequences \cite{Aalsma:2021qga, Cremonini:2020smy, Loges:2019jzs}. A particularly intriguing argument \cite{Cheung:2018cwt} is that the combination of EFT coefficients which appears in the WGC calculation of \cite{Kats:2006xp} is $(-1)$ times the same combination that appears in the shift to the black hole entropy at extremality. As a result, the mass shift will be negative, and the mild WGC satisfied, in any theory where the entropy corrections at constant charge and mass are positive at constant energy. This requirement was shown to hold for any UV completion which arises from integrating out states at tree-level, and conjectured to hold more generally \cite{Cheung:2018cwt}. We call this result the ``entropy-extremality relation", which schematically takes the form
\begin{align}
    \Delta m_{\text{ext}} \sim - T \, \Delta S \big|_{m = q}
\end{align}
A more general proof unrelated to black holes was given in \cite{Goon:2019faz}, and further details and examples were discussed in \cite{Hamada:2018dde, Cremonini:2019wdk, Cano:2019oma, Cano:2019ycn}.

Generically, the higher-derivative entropy gets first-order corrections from both the higher-derivative operators appearing in the Wald formula, and from the corrections to the horizon radius. From a geometric point of view, the entropy-extremality relation hinges on the fact that at $m = q$, the horizon shift corrections dominate the Wald-formula corrections. The corrections to the horizon radius also control the corrections to the mass shift, establishing the entropy-extremality relation in this limit. Later it was clarified \cite{Hamada:2018dde} that the horizon shift at $m = q$ is of order $\sqrt{c_i}$, where $c_i$ are the coefficients of the higher-derivative operators appear in the corrected action. The entropy-extremality relation is still valid because $m \sim c_i, \, S \sim \sqrt{c_i}$, and $T \sim \sqrt{c_i}$. These corrections to $T$ are inevitable if we hold the mass fixed because the relationship between $T$ and $m$ gets shifted and $m = q$ is no longer extremal. Thus in the corrected theory, $T$ is slightly positive at $m = q$-- specifically, it becomes order $\sqrt{c_i}$.

A common feature of the WGC and the positive entropy shift conjecture is that both depend on the mass. Relatedly, computing the mass shift and the entropy shift at fixed mass both require understanding of the boundary structure of spacetime. This makes them well-suited for studying spacetimes which are asymptotically flat or AdS, but for asymptotically dS spaces or other more cosmological solutions, it is not clear how they might be used as Swampland criteria (though see \cite{Montero:2019ekk} for a discussion of the WGC in dS). Given this, it may be interesting to try to understand how higher-derivative corrections affect quantities which are defined without reference to the boundary. Examples include any quantities which can be computed using only the near-horizon geometry.

The purpose of this note is calculate such a quantity: the entropy shift of 4d Reissner-Nordstr{\"o}m black holes at $T = 0$ (\textit{i.e.} at $m = q + \Delta m$) in the corrected theory. We do so using Sen's classical entropy function formalism \cite{Sen:2005wa}, which only makes reference to the near-horizon geometry. We then check this calculation using the explicit higher-derivative-corrected Reissner-Nordstr{\"o}m solution. We find that the $T = 0$ entropy shift is $\mathcal{O}(c_i)$, meaning that $
\Delta S$ is $\mathcal{O}(c_i)$ everywhere except for around $m = q$, where it is $\mathcal{O}(\sqrt{c_i})$. We also demonstrate this numerically. As a result, we find that the entropy shift at $T = 0$ is unrelated to the extremality bound. After showing that the $T = 0$ entropy shift can be written with a field-redefinition invariant set of four-derivative operators, we check it explicitly for the EFT that arises at low energies from QED with gravity, and we find that it is negative, in contrast to the fixed-$m$ entropy shift of \cite{Cheung:2018cwt}.

\section{4d Reissner-Nordstr{\"o}m Black Holes}

We will consider the following Lagrangian:
\begin{align}
\begin{split}
S &  = \int d^4 \sqrt{- g} \, \Bigg( \frac{R}{4} - \frac{1}{4} F^2 + c_1 R^2 + c_2 R_{\mu \nu} R^{\mu \nu} \\
& \quad + c_3 R_{\mu \nu \rho \sigma}R^{\mu \nu \rho \sigma} 
 \ + c_4 R F^2 + c_5 R_{\mu \nu} F^{\mu \rho} F^{\nu}{}_{\rho}\\
 & \quad +  c_6 R_{\mu \nu \rho \sigma} F^{\mu \nu} F^{\rho \sigma} + c_7 (F^2)^2  + c_8 F_{\mu \nu}F^{\nu \rho} F_{\rho \sigma}F^{\sigma \mu} \Bigg)  \, .
 \label{eq:lagrangian}
\end{split}
\end{align}

This theory admits solutions that are Riessner-Nordstr{\"o}m plus corrections, which are order $c_i$. Such solutions take the form
\begin{align}
\begin{split}
        ds^2 \ &= \ -f(r) dt^2 + \frac{dr^2}{g(r)} + r^2 d \Omega^2_{S^2} \, , \\
        F_{tr} \ &= \ \frac{q}{r^2} \, .
    \label{eq:RNshifted}  
\end{split}
\end{align}
where 
\begin{align}
\begin{split}
        f(r) = 1 + \frac{2 m}{r} + \frac{ q^2}{ r^2} + \Delta f(r) \, , \\
    \qquad g(r) = 1 + \frac{2 m}{r} + \frac{ q^2}{ r^2} + \Delta g(r) \, .
    \label{eq:RNfunsshifted}
\end{split}
\end{align}
In this paper, $\Delta$ will always denote a quantity which vanishes when $c_i \to 0$; such quantities are corrected with respect to the two-derivative solution with the same charge $q$. We can see the \textit{unshifted} black holes ($\Delta f = \Delta g = 0$) will be extremal when $m = q$, with horizon radius $r_h = q$. 

\section{Entropy Function}
\label{sec:ent function}

The classical entropy function is a method for computing the black hole entropy of extremal black holes \cite{Sen:2005wa} (see also \cite{Sen:2007qy} for a review). The method relies on the fact that extremal black holes have near-horizon geometries which factorize into AdS$_2 \times S^2$. The most general electrically charged such solution is
\begin{align}
\begin{split}
    ds^2 \ & = \ v_1 \left(-\rho^2 d\tau^2 + \frac{d\rho^2}{\rho^2} \right) + v_2 d\Omega^2 \, , \\
    F_{rt} \ &= \ e \, .
    \label{eq:nh_geom}
\end{split}
\end{align}
The idea, based on the logic of the attractor mechanism, is that the constants $v_1$ and $v_2$ are fixed by extremizing the entropy function, which equals the Wald entropy for the extremal RN black hole. The entropy function is constructed first by considering the function $f(v_1, v_2, e)$ defined by 
\begin{align}
    f(v_1, v_2, e) = \int d \theta d \phi \sqrt{-g} \mathcal{L} \, .
\end{align}
where $\mathcal{L}$ is the lagrangian density defined by (\ref{eq:lagrangian}). The entropy function is defined as the Legendre transform of this function, 
\begin{align}
    \mathcal{E} = 2 \pi \left( e q - f(v_1, v_2 e) \right) \, ,
\end{align}
where $e$ can be written in terms of $q$ using the condition
\begin{align}
    \frac{\partial f}{\partial e} = q \, .
\end{align}
The result for this procedure is 
\begin{align}
    \begin{split}
        e \ &= \ q \frac{v_1}{v_2} - \frac{4 q}{v_2} \Bigg( \left(2 - \frac{2 v_1}{v_2} \right) c_4 + c_5 + 2 c_6 \\
        & \qquad \qquad \qquad \qquad + \frac{4 q^2 v_1}{v_2^2} c_7 +  \frac{2 q^2 v_1}{v_2^2} c_8 \Bigg) \, . \\
        %
        %
        %
    \end{split}
\end{align}
Now it is a simple matter to extremize the entropy by demanding %
\begin{align}
    \frac{\partial \mathcal{E}}{\partial v_1} =0 \, , \quad \qquad \frac{\partial \mathcal{E}}{\partial v_2} =0 \, .
\end{align}
The corrected geometry corresponds to the values of $v_1$ and $v_2$ which extremize the entropy. We find
\begin{align}
\begin{split}
        & v_1 \ = \ q^2 +  8c_4+ 4 c_5 + 8 c_6 + 8 c_7 + 4 c_8 \, , \\
        & v_2 \ = \ q^2 + 8 c_4 - 8 c_7 -4 c_8 \, . 
\end{split}
\end{align}
The  extremal near-horizon geometry is easily recovered from the entropy functional formalism. When we plug our values of $v_1$ and $v_2$ back into the entropy function we obtain
\begin{align}
    \boxed{\mathcal{E} = \pi q^2 - 4 \pi \left( 2 c_2 + 4 c_3 + c_5 + 2 c_6 + 2 c_7+ c_8 \right) \, .}
    \label{eq: entropy function result}
\end{align}

This is the entropy of the extremal black hole-- we see that it is given by $ \mathcal{E} = A / 4 + \Delta S$. 

\section{Extremality and Entropy from the Geometry Shift}
\label{sec:geom}

To confirm and to better understand this result, let us consider the geometry shift. In \cite{Kats:2006xp, Cheung:2018cwt}, it was shown that this takes the form
\begin{align}
\begin{split}
    & \ \ \Delta g(r) \ = \ -\frac{4 q^2}{5 r^6} \Bigg( ( c_2+ 4 c_3) (12 q^2 - 30  m r + 20 r^2 ) \\
    & + c_4 (30 q^2 - 70 m r + 40 r^2) + c_5 (11 q^2 - 25 m r + 15 r^2)  \\ 
    & \qquad + c_6 (16 q^2 - 35 m r + 20 r^2) + c_7  (2q^2) + c_8  (q^2)  \Bigg) \, .
    \label{eq:KMPlag}
\end{split}
\end{align}

\subsection{Extremality Shift}

First we compute the shift to extremality, which is defined as the mass shift at zero temperature, again holding $q$ fixed throughout the calculation. The extremal mass is $m_{\text{ext}} = q + \Delta m$. At this value of the mass, there is also a shift to the horizon radius, $r_h = q + \Delta r$. 

If we work exactly at extremality, $m = m_{\text{ext}}$, then $g(r_h) = g'(r_h) = 0$. This gives us two conditions, which we can solve for $\Delta m$ and $\Delta r$ (see the appendix for the expression in terms of $g$). The result is 
\begin{align}
\begin{split}
    \Delta m \ &= \ -\frac{2}{5 q} \left( 2 c_2 + 8 c_3 + c_5 + c_6 + 2 c_7 + c_8 \right) \, , \\
    \Delta r \ &= \ \frac{2}{q} \left(2 c_4 - 2 c_7 - c_8 \right) \, .
    \label{eq: mass shift result}
\end{split}
\end{align}
The mass shift here is the result of \cite{Kats:2006xp}. The radius shift is known to be related to the mass shift away from extremality and at unshifted extremality $m = q$ \cite{Cheung:2018cwt, Hamada:2018dde}. However we see that the radius shift is not proportional to $\Delta m$ at $T = 0$. 

\subsection{Entropy Shift}

The Wald entropy formula \cite{Wald:1993nt} allows us to compute the entropy of higher-derivative-corrected black holes.
%
%
For spherically symmetric spacetimes, this takes the simplified form
\begin{align}
    S = - 2 \pi A \frac{\delta \mathcal{L}}{\delta R_{\mu \nu \rho \sigma}} \epsilon_{\mu \nu} \epsilon_{\rho \sigma} \Big|_{\text{hor}} \, .
\end{align}
Higher-derivative corrections contribute in two ways: first, they shift the radius, and therefore $A$. Second, the interactions themselves introduce Riemann tensors into the Lagrangian which must be differentiated. We will call these two contributions $\Delta S_h$ and $\Delta S_i$ respectively. How they are computed is nicely explained in \cite{Cheung:2018cwt}, so we will not review the details here. The results are
\begin{align}
\begin{split}
    \Delta S_h \ &= \ \frac{\Delta A}{4}  = 2 \pi r \Delta r \ = \ 4 \pi \,  (2c_4 - 2c_7 - c_8) \, , \\
    \Delta S_i \ &= \ -4\pi \, (2 c_2 + 4 c_3 + 2 c_4 + c_5 + 2 c_6 ) \, .
\end{split}
\end{align}
The total entropy is the sum of both of these
\begin{align}
    \Delta S = - 4 \pi (2 c_2 + 4 c_3 + c_5 + 2 c_6 + 2 c_7 + c_8 ) \, ,
    \label{eq:entshift}
\end{align}
in agreement with the result of the entropy function \eqref{eq: entropy function result}. This expression is not proportional to the mass shift \eqref{eq: mass shift result}.

\subsection{Numerical Test}

The geometry shift given above actually allows us to probe what happens for any value of the mass, not just at zero temperature. In particular, we should be able to see that the horizon shift is $\mathcal{O}(c_i)$ at all points except for at $m = q$. To do this, let us choose $c_7 = 10^{-4}$ with all other $c_i = 0$. Then we can study the behavior of $g(r)$ numerically. 

\begin{figure}[t]%
    \centering
    \subfloat[\centering]{{\includegraphics[width=8cm]{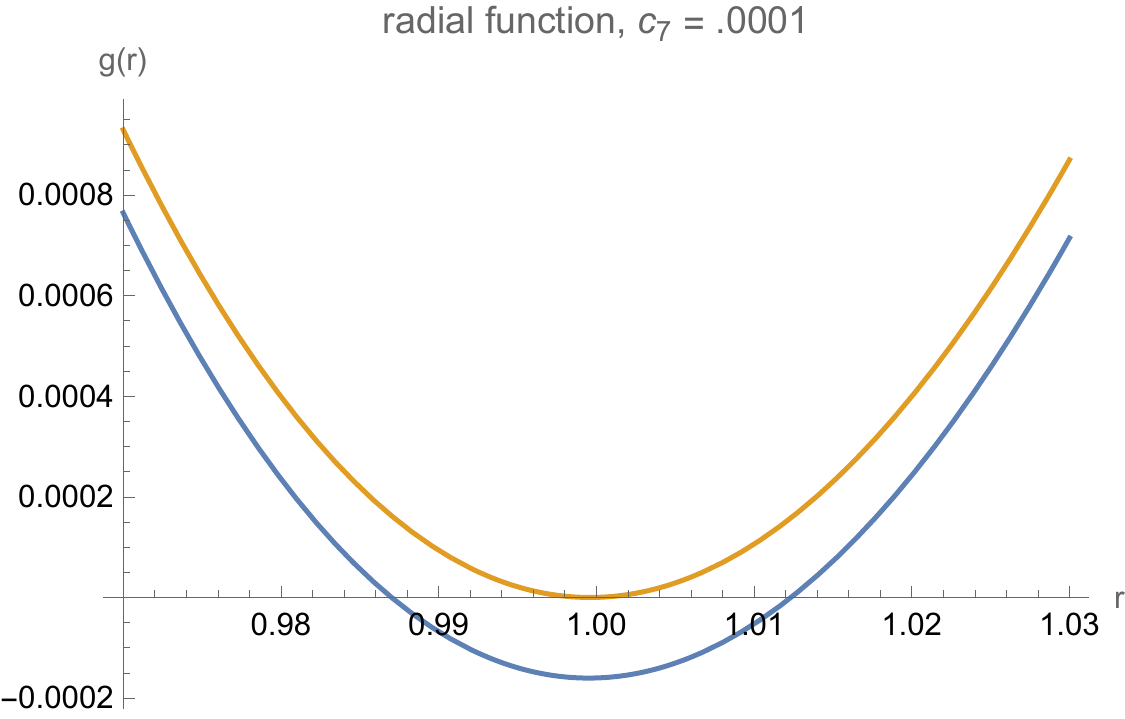} }}%
    \qquad
    \subfloat[\centering ]{{\includegraphics[width=8cm]{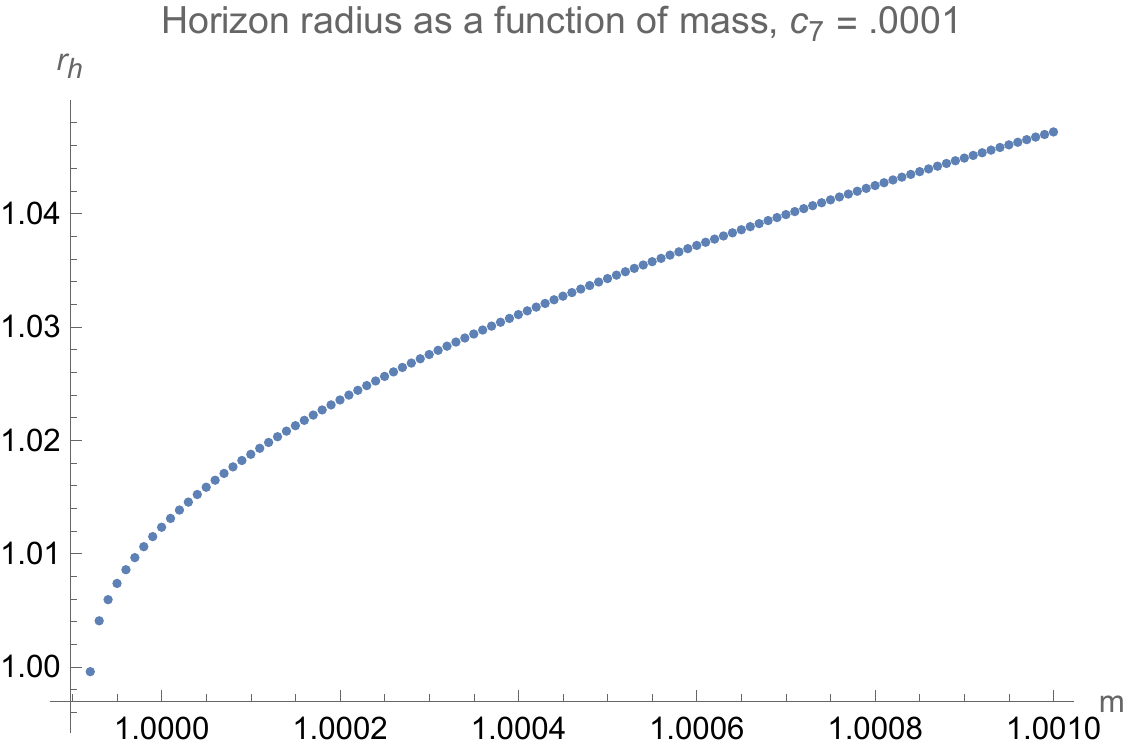} }}%
    \caption{(a) Radial functions with only $c_7$. Blue is unshifted, orange shifted extremality. (b) Horizon shift with only $c_7$, for different values of the mass.}%
    \label{fig1}%
\end{figure}

At $m = q$ we find $\Delta r  = \pm .0125 \sim \sqrt{c_7}$ but at $m = 1 + \Delta m$, $\Delta r = -.0004 \approx -4  c_7$. Furthermore we can plot the horizon radius as a function of mass, as we have done in figure~\ref{fig1}(b). This shows us that the horizon radius varies smoothly down to its extremal value, and dips slightly into the negative region very near the true value for the extremal mass. 

\section{Positivity}

To draw any physical conclusions from this calculation, we must translate into a basis which is invariant under field redefinitions. We use the basis $\{d_0, d_3, d_6, d_9 \}$ of \cite{Cheung:2018cwt}. The mass and entropy shifts become
\begin{align}
    \Delta m = -\frac{2}{5 q} d_0, \quad \Delta S = -4 \pi (d_0-2d_3+d_6)
    \label{eq:field redefs}
\end{align}
The horizon shift we obtained cannot be written in this form. This is a little strange but ultimately irrelevant for our purposes. It is interesting that $d_0$, the combination which is related to the WGC, enters the $T = 0$ entropy shift with the opposite sign as the constant $m$ entropy shift of \cite{Cheung:2018cwt}. Furthermore, consider that for magnetic black holes, $\Delta m \sim - (d_0 + 2 d_6)$ \cite{Jones:2019nev} (see appendix also). Therefore enforcing the WGC (or alternatively, the positivity of the fixed-mass entropy shift) for all charged black holes in 4d tells us
\begin{align}
    d_0+ d_6>0 \, .
    \label{eq:d0d6bound}
\end{align}

The coefficient $d_3$ multiplies $R_{\mu \nu \rho \sigma} R^{\mu \nu \rho \sigma}$ in the action. We see that if $d_3>0$ then the extremal entropy shift can have either sign. But if $d_3<0$, then the extremal entropy shift must be negative to be consistent with the mild WGC.  (It is also interesting to note that in AdS, the positivity of the fixed-mass entropy implies that the coefficient of Riemann squared is positive \cite{Cremonini:2019wdk}-- the argument does not extend to flat space because there are no neutral black holes which are thermodynamically stable.)

It is tempting to try to relate this to a statement about the Gauss-Bonnet term, which can be related to $d_3$ by field redefinitions. This term gives a topological contribution to the Wald entropy that appears (with either sign) to allow for violations of the 2nd law thermodynamics during black hole mergers or formation \cite{Jacobson:1993xs, Liko:2007vi, Sarkar:2010xp}. However, the present case (\textit{i.e.} involving the full set of coefficients) is more complicated. Even if the Gauss-Bonnet term is zero, we can still have a $d_3$ contribution, which arises after breaking $d_3$ into a part which depends on Gauss-Bonnet and a part which does not. See the appendix for details on field redefinition invariance and Gauss-Bonnet invariance.

\subsection{QED with Gravity} 

The purpose of this work has been to consider the entropy shift at $T = 0$, and to explore the extent to which this might be an interesting quantity for bounding higher-derivative corrections. We find that this shift can also be written in a field-redefinition-invariant basis, which is a basic requirement on physically meaningful quantities. However it is not related to the other shifts, and there does not appear to be anything inconsistent with its negativity \footnote{Note also that the proof \cite{Cheung:2018cwt} that the constant-mass entropy shift is positive for tree-level completions does not apply to this case.}. For an explicit example, one can consider QED with gravity, and compute the EFT coefficients that arise from integrating out the electron. These corrections were computed in \cite{PhysRevD.22.343}\footnote{see also \cite{Cheung:2014ega, Goon:2016une, Aalsma:2020duv} for discussions in the context of the WGC and other positivity bounds}:
\begin{align}
\begin{split}
        & c_7 = - \frac{5 \alpha^2}{180 m_e^4}\,, \qquad  c_8 = \frac{14 \alpha^2}{180 m_e^4} \\
        & c_4 = \frac{\alpha}{144 \pi m_e^2}, \quad c_5 = -\frac{13 \alpha}{360 \pi m_e^2}, \quad  c_6 = \frac{\alpha}{360 \pi m_e^2} \, .
\end{split}
\end{align}
Our conventions imply $M_p = 1/\sqrt 2$, so we can restore $M_p$ in the ratios to find
\begin{align}
    \Delta S|_{T = 0} =  -4 \pi \left( \frac{16\alpha^2}{180} \frac{M_p^4}{m_e^4} - \frac{22 \alpha}{360 \pi} \frac{M_p^2}{m_e^2} \right) 
\end{align}
So the extremal entropy shift must be negative because $\alpha \gg (M_p/m_e)^2$. This example is particularly interesting because it is likely very close to our universe since the electron is the lightest charged particle. However this has not been confirmed through measurement, and it is possible that other sources change the answer. 

Part of the motivation of this is to provide some insight on the entropy-extremality relation, which relates the mass shift to the entropy shift at fixed mass. From a geometric point of view, the shift to entropy at constant mass is special because it is controlled by the geometry shift $\Delta g(r)$. That means, in particular, that it cannot be affected by topological contributions to the entropy, which are possible in principle but which can't have any local effects, such as contributions to solutions to the equations of motion such as $g(r)$. Such topological corrections can, in principle, be addressed by the $T = 0$ entropy shift.

Another motivation here is that both the WGC and the positive entropy conjecture of \cite{Cheung:2018cwt} are limited by the fact that they depend on the mass. In particular, this makes it difficult to understand what, if any, form should hold in de Sitter space \footnote{nonetheless, the WGC was previously considered in dS in \cite{Montero:2019ekk}}, where there is no positive conserved energy, rendering the mass somewhat ambiguous. It would be interesting to try to extend the ideas discussed in this paper to de Sitter space. This also raises the possibility that some form of these positivity conjectures might be relevant for the cosmological horizon in addition to black hole horizons. 

Another possible direction would to compare the constant-mass and constant-temperature entropy shifts in more concrete examples (for instance by computing them using the higher-derivative coefficients computed for UV complete holographic theories in \cite{Bobev:2021oku}). It would also be interesting to continue to investigate what can be learned from the near-horizon limit. In this work we have only computed the classical entropy, but in principle higher-derivative corrections to the \textit{quantum} entropy \cite{Sen:2008vm} could be addressed as well, perhaps along the lines of recent work such as \cite{Iliesiu:2020qvm, Heydeman:2020hhw, Banerjee:2021vjy}.


\section*{Acknowledgments}
The author thanks Callum Jones for an interesting conversation which led to this project. Also, thanks to Gary Shiu, Toshifumi Noumi, Grant Remmen, Sera Cremonini, and Jim Liu for insightful comments and feedback on this paper. This work was supported by the European Research Council (ERC) under the European Union's Horizon 2020 research and innovation programme (grant agreement no.~758903).

\appendix

\section*{}
\bibliography{cite.bib}

\section{Horizon Shift from Geometry Shift}

Here we will be more explicit about how to compute the horizon shift using the shifted geometry. This should make it more clear why our procedure applies to ``shifted extremality", $m_{\text{ext}} = q + \Delta m$. In our notation, $g = g_0 + \Delta g$, $r_h = r_0 + \Delta r = q + \Delta r$. For shorthand we will use $\partial_rg(q) \equiv \partial_r g(r) |_{r = q}$, etc.

\subsection{Away from extremality}

When the temperature is large compared to the value of the EFT coefficients, it is simple to extract the horizon shift:

\begin{align}
\begin{split}
    & g(r_h) = g_0(q) + \Delta r \, \partial_r g_0(q) + \Delta g(q) \\
    & \qquad \implies \Delta r_h = -\frac{\Delta g(q)}{\partial_r g_0(q)}
\end{split}
\end{align}
This formula is correct except for at extremality, where $\partial_r g_0(q)$ goes to zero and the horizon shift blows up, as pointed out in \cite{Cheung:2018cwt}. This indicates that a different procedure is needed to describe the horizon shift for extremal black holes.

\subsection{Horizon shift at $m = q$}

A method to compute the entropy at $m = q$ was given in \cite{Hamada:2018dde}. Because $\partial_r g_0(q) \to 0$ at extremality, we need to look at the second order of the series around $r = q + \Delta r$. So first we expand
\begin{align}
\begin{split}
    g(r_h) \ &= \ g_0(q) + \Delta g(q) \\
    & \quad  + \Delta r \, \partial_r g_0(q) + \frac{1}{2} (\Delta r)^2 \, \partial_r^2 g_0(q) \, .
    \label{eq:Hamada_ext}
\end{split}
\end{align}
As we approach extremality, $g_0(q)$ and $\partial_r g_0(q)$ go to zero. The result is
\begin{align}
     (\Delta r)^2 = -\frac{1}{2} \frac{\Delta g(q)}{ \partial_r^2 g_0(q)}
\end{align}

This calculation cannot be extended to shifted extremality, $m_\text{ext} = q + \Delta m$ because $g_0(q)$ does not vanish-- actually $g_0(q) = \Delta m \, \partial_m \, g_0(q)$. We can also easily compute the temperature shift in this limit
\begin{align}
    T = \frac{1}{4 \pi} \sqrt{ g'(q + \Delta r) f'(q + \Delta r)} = \frac{1}{4 \pi} \Delta r g''_0(q) \, .
\end{align}
Again we see that everything in the $m \to q$ limit is controlled by $\Delta r$. If we go away from this limit, then we need to consider also the corrections to $f(r)$, which generally are different from the corrections to $g(r)$ (however both functions must have the same zeros for consistency of the metric signature).

\subsection{Horizon shift at $T = 0$}
\label{sec:method at ext}

For truly extremal black holes, the correct procedure is to impose extremality first, and then solve for $\Delta m$ and $\Delta r$ at the same time. We first impose
\begin{align}
    \begin{cases}
      g(q + \Delta m, q + \Delta r) \ &= \ 0\\
      \partial_r g(q + \Delta m, q + \Delta r) \ &= \ 0
    \end{cases}\,.
\end{align}
Expanding these two equations around $m = q$ and $r = q$ leads to
\begin{align}
    \begin{cases}
      \Delta m \, \partial_m g_0(q, q) + \Delta g(q,q) \ &= \ 0\\
      \Delta r \,  \partial_r^2 g_0(q , q ) + \Delta m \,  \partial_m \partial_r g_0(q , q ) + \partial_r \Delta g(q,q) \ &= \ 0
    \end{cases}\,.
\end{align}
We can solve this system, obtaining
\begin{align}
\begin{split}
    \Delta m \ &= \ - \frac{ \Delta g(q,q)}{ \partial_m g_0(q, q)} \, , \\
    \qquad \Delta r \  &= \  \frac{ \Delta g(q,q) \,  \partial_m \partial_r g_0(q , q )- \partial_r \Delta g(q,q) \partial_m g_0(q, q)  }{\partial_r^2 g_0(q , q ) \partial_m g_0(q, q)} \, .
\end{split}
\end{align}
These formulas, combined with the geometry shift, can be used to reproduce the results for the mass shift and horizon shift in the body of the paper. In particular, we see that $\Delta m$ and $\Delta r$ are both first-order in $c_i$.

\section{Shifted near-horizon geometry}

A useful check is to use our shifted RN geometry \eqref{eq:KMPlag} to reproduce the near-horizon geometry we found by maximizing the entropy function. This is accomplished by replacing 
\begin{align}
    t \to \tau \frac{r_h^2}{\lambda}, \qquad r \to \lambda \rho + r_h \, ,
\end{align}
and then taking $\lambda \to 0$. This procedure maps the shifted RN solution \eqref{eq:RNshifted} to 
\begin{align}
    ds^2 = -\frac{1}{2} f''(r_h) \rho^2 d\tau^2 + \frac{2}{g''(r_h)} \frac{d \rho^2}{\rho^2} + r_h^2 d\Omega^2 \, .
\end{align}
From this we can read off the near horizon geometry. We ignore the $d\tau^2$ term, which can be changed by rescaling $\tau$ or $\rho$, unlike the $d \rho^2$ term. Therefore it is the $d \rho^2$ term which fixes $v_1$:
\begin{align}
    v_2 = \frac{2}{g''(r_h)} = q^2 + 4 (2 c_4 + c_5 + 2 c_6 + 2 c_7) \, ,
\end{align}
while $v_2$ is simply computed by 
\begin{align}
    v_2 = r_h^2 = q^2 + 2 \Delta r = q^2 + 8 (c_4 - c_7) \, .
\end{align}
This confirms that the entropy function calculation gives the correct near-horizon geometry.

\section{Generalization to dyonic black holes}

It is also interesting to inspect the structure of this calculation for the full set of charged black holes in 4d flat space-- namely, the dyonic black holes. In this case the near-horizon geometry takes the same form as \eqref{eq:nh_geom} except we add $F_{\theta \phi} = p \sin \theta$. Repeating the extremization procedure gives us the corresponding results for the geometry
\begin{align}
\begin{split}
    v_1 \  = \ q^2 + p^2 + \Big(& 8 \frac{q^2-p^2}{q^2 + p^2}c_4+ 4 \frac{q^2}{q^2 + p^2}c_5 + 8 \frac{q^2}{q^2 + p^2}c_6 \\
    & \qquad + 8 \frac{(q^2-p^2)^2}{(q^2 + p^2)^2} c_7+ 4 \frac{q^4+p^4}{(q^2 + p^2)^2} c_8 \Big)\\
    v_2 \  = \ q^2 + p^2 + \Big(& - 8 \frac{p^2-q^2}{q^2 + p^2}c_4- 4 \frac{p^2}{q^2 + p^2}c_5 - 8 \frac{p^2}{q^2 + p^2}c_6 \\
    & \qquad - 8 \frac{(p^2-q^2)^2}{(q^2 + p^2)^2} c_7- 4 \frac{p^4+q^4}{(q^2 + p^2)^2} c_8 \Big)
\end{split}
\end{align}
In this case, the shifted near-horizon geometry takes a much more symmetric form; the shifts to $v_1$ and $v_2$ are related by swapping $q$ and $p$, and changing sign. From this we can extract the shifted entropy:
\begin{align}
\begin{split}
    \Delta \mathcal{E} \ = \ - 4 \pi \Bigg(& 2 c_2 + 4 c_3 + c_5 + 2 c_6 \\
    & + 2 \frac{(q^2-p^2)^2}{(p^2 + q)^2} c_7 + 2 \frac{q^4+p^4}{(p^2 + q)^2} c_8  \Bigg)
\end{split}
\end{align}

The mass shift cannot be extracted from the near-horizon geometry, and so the calculation from the full geometry must be performed for dyonic black holes \cite{Jones:2019nev}. The result is 
\begin{align}
    \begin{split}
        \Delta m = -\frac{2}{5 \sqrt{q^2 + p^2}} \Bigg(& 2 c_2 + 8 c_3 + c_5 + \frac{q^2 + 3 p^2}{p^2 + q^2} c_6 \\
        & + \frac{2 (q^2 - p^2)^2}{(p^2 + q^2)^2} c_7+ \frac{q^4 +  p^4}{(p^2 + q^2)^2} c_8 \Bigg)
    \end{split}
\end{align}
Using Mathematica's Reduce function, it is possible to see that $\Delta \mathcal{E}>0$ and $\Delta m < 0$ are consistent, meaning there are choices of the coefficients which satisfy both constraints for black holes of any possible charge. It would not be difficult to map the space of coefficients allowed if we require $\Delta S > 0$ or $\Delta S < 0$, but we will not pursue this at this time.

\section{Field-redefinition invariance and Gauss-Bonnet independence}

Above we discussed that all local quantities must not depend on the Gauss-Bonnet coefficient. Here we would like to clarify that point. First, consider the following basis of field-redefinition-invariant operators \cite{Cheung:2018cwt}:
\begin{align}
    \begin{split}
        d_0 \ & = \ 2 c_2 + 8 c_3 + c_5 + c_6 + 2 c_7 +  c_8  \, ,  \\
        d_3 \ & = \ 2 c_3 \, ,  \\
        d_6 \ & = \ c_6 \, , \\
        d_9 \ & = \ 2 c_2 + c_5 + \frac{1}{2} c_8 \, . \\
    \end{split}
\end{align}
Any term which is determined by \textit{local} physics, such as the solution to the equations of motion, must not change if we add the Gauss-Bonnet combination
\begin{align}
    G = R_{\mu \nu \rho \sigma} R^{\mu \nu \rho \sigma} - 4 R_{\mu \nu } R^{\mu \nu} + R^2 \, .
\end{align}
to the lagrangian. Now, in this paper, $R^2$ has no effect because $R = 0$ on our background, so we require local quantities to be invariant under $c_2 \to c_2 - 4 \alpha$ and $c_3 \to c_3 + \alpha$.

It is clear that $d_0$ already has this invariance, as it should since it is the WGC combination which is controlled by the geometry shift $\Delta g$. In fact, for dyonic black holes with $p = q$, $\Delta m \sim 4 d_3 + 2 d_6 + d_9$. Therefore we see that this shift is also independent of the Gauss-Bonnet coefficient even though it depends on $d_3$. For the bare $d_3$ term, it is best to transform it into a term which includes Gauss-Bonnet and one which is independent of it. This is done using
\begin{align}
    d_3 \to -\frac{1}{15} (d_3 - 4 d_9) + \frac{4}{15}(4d_3 + d_9) \, .
\end{align}
The first term depends on Gauss-Bonnet and the second term is independent of it. So we see that requiring independence from Gauss-Bonnet is not enough to imply $d_3 = 0$. It only implies $d_3 = 4 d_9$.

\end{document}